# Surface Reading LLMs:
# Synthetic Text and Its Styles

Hannes Bajohr
University of California, Berkeley
hbajohr@berkeley.edu

## ABSTRACT

Despite a potential plateau in ML advancement, the societal impact of large language models lies not in approaching superintelligence but in generating text surfaces indistinguishable from human writing. While Critical AI Studies provides essential material and socio-technical critique, it risks overlooking how LLMs phenomenologically reshape meaning-making. This paper proposes a semiotics of "surface integrity" as attending to the immediate plane where LLMs inscribe themselves into human communication. I distinguish three knowledge interests in ML research (*epistemology*, *epistēmē*, and *epistemics*) and argue for integrating surface-level stylistic analysis alongside depth-oriented critique. Through two case studies examining stylistic markers of synthetic text, I argue how attending to style as a semiotic phenomenon reveals LLMs as cultural machines that transform the conditions of meaning emergence and circulation in contemporary discourse, independent of questions about machine consciousness.

## 1 INTRODUCTION

With the release of GPT-5 in August 2025, little remained of the confidence OpenAI CEO Sam Altman had projected a year earlier, when he touted it as a milestone on the road to artificial superintelligence (NBC News 2024). The model was better, but not by much—certainly no leap like GPT-4 had been over GPT-3. As OpenAI's stock slipped, talk quickly turned to a plateau in machine learning (ML) research (Wilkins 2025),[1] muting both the hopes and fears of imminent self-improving systems that had long preoccupied the industry (Amodei 2024; Kokotajlo et al. 2025). Whether this will be a temporary snag or the start of the hype cycle's decline, however, matters little. The real watershed had already come three years earlier with the public rollout of easily accessible large language models. Altman's promised revolution was long here, not because this was the road to simulated consciousness or superintelligence, but because anyone could now make use of machines that produce writing virtually indistinguishable from human text. The more AI writing intervenes in our own, slotting itself into our communication, and becoming part of the greater epiphylogenesis of human technics, the clearer it is what is truly new here is something else—a novel *semiotic* state. Consequently, calls for a semiotic analysis that rises to this situation have been mounting.

In the altered "textual condition" (McGann 1991) that is probabilistic ML, developing a semiotics of LLMs, however, requires a delicate balancing act: attending to the immediacy of the sign as it appears to us while recognizing its dependence on the techno-social contexts of production and circulation. Current critical writing on ML remains largely focused on the latter, valuing depth over surface. While such approaches are indispensable, they are also incomplete and risk overlooking the very phenomenon that makes LLMs semiotically interesting: their capacity to generate surfaces that exhibit meaning-effects in their own right. Acknowledging the important work that is done in Critical AI Studies to engage with the intellectual, material and historical reality of ML systems, the aim of this essay is to make a case for integrating a phenomenological approach to their outputs into this field as well—an approach that honors what I call their "surface integrity."

In doing so, I offer an appeal not to be limited in one's research by the rhetoric of depth and the calls for the priority of foundational insights over

The first seed for this essay was a talk I presented at the opening of the *Center for the Humanities and Machine Learning* at the University of California, Santa Barbara, in January 2025. I want to thank Rita Raley and Fabian Offert as well as the attendants for helpful feedback. I also want to thank Stephen Best, Shane Denson, Noam Elcott, Jules Pelta Feldman, Bernard Geoghegan (as well as the attendants of his AI reading group), Marti Hearst, Alan Liu, Sarah Pourciau, Sonja Thiel, Christina Vagt, and the members of the DFG-Netzwerk "Das Wissen der digitalen Literatur" for their suggestions.

[1] I shall use the term "machine learning" instead of "artificial intelligence" throughout, unless the discourse has established the latter, as e.g. in the case of "Critical AI Studies."

ostensibly superficial evaluation. It is true that the surface often serves as a screen, both for projecting hopes and fears and for shielding curious eyes from whatever lies behind. But this surface is also the principal plane of engagement with the phenomena of machine learning—the plane on which *we encounter them in the life-world*. We do well not to lose sight of this plane and to take it seriously before discarding it too quickly as a mere epiphenomenon of what is "really going on underneath." That this is by no means trivial becomes clear as we consider that reading these surfaces comes with problems that other texts, or even other technical systems, did not yet pose. I suggest that the way into this problem and toward a practice of developing a technological imagination as probabilistic intuition that passes through the venerable concept of "style," understood not as *ornatus*, but as the key to understanding how these semiotic systems inscribe themselves into culture. Only by attending to the surface can we grasp how LLMs reshape the very conditions under which meaning emerges and circulates, long before machines have become intelligent.

## 2 QUESTIONS OF METHOD

Before I "delve into" (as ChatGPT would say) (Kobak et al. 2025) what I mean by "surface integrity" and provide two brief examples of how reading its styles might look, I want to start by unpacking the epistemic and methodological relationship between the humanities and machine learning. In doing so, I also engage with recent discussions about the presence—or absence—of a methodology of Critical AI Studies and the mode of critique they presuppose.[2]

This requires acknowledging a key challenge: it makes little sense to speak of "the humanities" as a monolith, since the disciplines vary significantly in how they are impacted by and respond to the proliferation of machine learning technologies (Kirschenbaum and Raley 2024). These differences are fundamental because they shape the kinds of epistemic inquiry and methodology that each field is engaged in, but also what they, in concert, can achieve with ML as their shared "boundary object." (Star 1989). In fact, a certain limitation of that approach—what leads me to stress the value of surfaces—is central to my argument.

By now, even some STEM researchers recognize that the humanities may be useful in examining the impact of ML, even if it all-too often remains confined to an idea of "ethics" in a largely justificatory role (Munn 2023). Yet as soon as the technical disciplines are confronted with soft and fuzzy categories like tone, affect, narrative, or, in the widest sense, *culture*, the humanities—as what Ernst Cassirer has called *Kulturwissenschaften*, the "sciences of culture"—appear to have a role to play (Cassirer 1961). The precise nature of this role, however, remains contested and is largely shaped by disciplinary perspectives. This is not least due to the fact that humanistic fields bring divergent "knowledge interests" to their objects, to abuse Jürgen Habermas's term, which has an effect on the methodologies they employ.

Very broadly speaking, we can distinguish three of these knowledge interests when it comes to the study of ML systems that I will term the *epistemology*, the *epistēmē*, and the *epistemics* of machine learning. These categories are, of course, ideal types painted with the broadest of brushes, and they inevitably blur into one another, but they nonetheless serve as a useful heuristic for gaining some orientation in the field. (See Dhaliwal (2023) for an alternative model.)

The first knowledge interest, the *epistemology* of machine learning, aligns closely with the traditional philosophical study of knowledge as it pertains to technical artifacts. This perspective raises foundational questions about intelligence, intent and intentionality, agency, and meaning as they manifest—or fail to manifest—in ML systems in general. At its core, this approach grapples with whether these systems can genuinely possess or simulate cognitive capacities, such as understanding or reasoning, or whether they merely exhibit patterns that mimic such capacities. It also explores the ethical implications of agency and responsibility, asking whether ML systems can or should be considered moral agents—or if responsibility rests solely on the designers, users, or societal structures that deploy them. Here, we would find figures like, in the past, Hubert Dreyfus (1992) or John Haugeland (1989), or, in the present, Brian Cantwell Smith (2019) or Margaret Boden (2016). While primarily a philosophical endeavor, the epistemology of ML can be operative secondarily in other fields, for instance in literary studies and linguistics (particularly if intent is understood to be a prerequisite of authorship or communication in general) or anthropology and social theory (where questions of agency and structure are central)—but it can also inform actual AI development: I may call to mind the research program of a "Heideggerian AI" in the late 1980s and early 1990s inspired by Dreyfus's writings.[3] The epistemology of ML, then, can extend to any discipline where the discourse on machine learning depends on the theorization of its actual capacity—or incapacity—to function as direct or indirect cognizer, agent, or mediator of meaning and action.

[2] The following offers just a partial view of recent developments in Critical AI Studies as a distinct field: Goodlad (2023), Kirschenbaum (2024), Offert (2023), Offert and Dhaliwal (2024), and Raley and Rhee (2023).

[3] The main ones are Agre (1997), Brooks (1991), and Winograd and Flores (1986). Dreyfus responded to these attempts with broad rejection, see (Dreyfus 2007).

Second, we might turn to the *epistēmē* of machine learning, which invites an exploration of the historically contingent, culturally specific, and often unexamined theoretical assumptions underpinning the design of machine learning systems. This approach aligns with a Foucauldian project of uncovering the historical a prioris that shape the broader enterprise of artificial intelligence research and tracing its genealogy. Here, the focus shifts from the inherent epistemic abilities ML systems do or do not display to the intellectual and cultural conditions that enable, steer, and constrain their development. It interrogates how particular historical knowledges, philosophical paradigms, and cultural values have informed the conceptualization and implementation of AI. For this approach, the reductionist tendencies of Cartesian rationalism, the quantification of human experience in 19$^{th}$ century positivism, or the cybernetic theories of the mid-20$^{th}$ century have all left their mark on how ML is a technical embodiment of dispositifs of knowledge. Here, we might think of Katherine Hayles (1999), Matteo Pasquinelli (2023), Adrian Mackenzie (2017), or Orit Halpern (2014) who, in one way or other, follow Philip Agre's observation that "artificial intelligence is philosophy underneath" (Agre 1995, 3) and aim to make this implicit philosophy explicit.

The third knowledge interest is concerned with what I want to call the *epistemics* of ML systems. By this I mean the applied, practical, and material examination of *specific, actually existing* technical systems in their contexts. This is the largest and messiest field, and it includes everything before, during, and after the deployment of systems—from the political economy of tech development to its concomitant extractivism of earth, data, and labor, to the functionality of the employed systems themselves. The critique of these factors highlights, on the one hand, the ways in which the signatures of control—political, financial, ideological, ecological—are inextricably entwined with ML. On the other, it discusses the ways in which these factors are integrated in the very material structure of these systems—in the data sets, the training algorithms, and the pipelines of their deployment. This domain is primarily populated by scholars from media and technology studies who have long developed a robust repertoire of techno-critical frameworks; I also believe that it is here that we would find many, though certainly not all, of the researchers converging now under the banner of "Critical AI Studies," who have illuminated the entanglements of machine learning systems with systemic inequalities, economic and radicalized structures, and ecological disaster; Kate Crawford (2021), Justin Joque (2022), Wendy Chun (2021), Ruha Benjamin (2019), and Louise Amoore (2020) are only a few names that stand for this strand, but it should also include scholars that have combined computational methodologies with critical cultural analysis, such as Deb Raji et al. (2020), Julia Angwin et al. (2016), Joy Buolamwini, or Timnit Gebru (Gebru and Buolamwini 2018), who have shown how empirical studies of data sets, algorithms, and their societal applications can expose the power structures, ideologies, and production of reality inherent in these systems.

This, then, is the triple knowledge interest I see in the field: The epistemology of machine learning examines, and sometimes debunks, its cognitive claims; the epistēmē of machine learning uncovers its often hidden historical, conceptual, and cultural assumptions; and the epistemics of machine learning offers a material critique of its underlying socio-technical systems. While Critical AI Studies might be mostly at home in epistemics, we certainly can find branches of it in all three knowledge interests regarding machine learning.

Yet while culture may be the common, if sometimes overly capacious, concern of the humanities, and machine learning a potential boundary object through which that concern now takes shape, there is no reason to expect a single methodology. Each field brings its own toolkit, ranging from conceptual analysis, media archaeology, discourse analysis, critical code studies, and STS-inflected ethnography to critical legal studies and the many methods developed within the digital humanities and adjacent fields. Even within a shared knowledge interest, then, there exists a wide spectrum of approaches. For this reason, Critical AI Studies in particular is currently marked by debates over method—whether to integrate empirical techniques more fully, how specific discussions of models and architectures should be, and whether the field risks becoming ancillary to, and complicit with, the development of ethically and politically dubious ML systems (Goodlad 2023; Lindgren 2024; McQuillan 2022; Raley and Rhee 2023; Roberge and Castelle 2021).

Increasingly, we also find meta-methodical concerns. The debate is no longer only about the whether or the how of ML, but about the wherefore. A recurring question is that of the *audience* of ML critiques, since more often than not, such studies risk preaching to the choir of those who are already convinced that technology is, first and foremost, a vehicle of the exercise and extension of power. Does much actually follow from repeating this point ad nauseam, only with different technical objects? The criticism can escalate into the polemic that the result of the analysis is already clear, only the way to get there needs to be found. The reality, of course, is more nuanced. But these tensions highlight the fact that the field is wrestling with the status and purpose of its very criticality. David Bates, in his genealogy of the constructedness of human intelligence, means exactly this when he writes about AI critique: "We (critical humanists) always know *in advance* that the historical unraveling will reveal, say, the importance of money, or political and institutional support, or exclusions in establishing what is always contingent." (Bates 2024, 6)

Ranjodh Singh Dhaliwal, too, articulates this problem in his review of recent work in Critical AI Studies, observing that technology critique has become trapped in predictable patterns where scholars "always, as Bruno Latour stated in his screed 'Why Has Critique Run Out of Steam?,' [are] 'debunk[ing] objects [that the critics] don't believe in by showing the productive and projective forces of people.' " For Dhaliwal, this creates a fundamental question of purpose: "Who, one may ask, needs to know that this or that technology is not neutral? Is it the layperson, the technoscientific disciplinarian, or perhaps the critic more than anyone else?" His diagnosis is particularly damning when he suggests that "we cannot make technology critiques about much" and that critics are "screen watching analytic reruns of sociotechnical priors" (Dhaliwal 2023, 312, 316).

Connected with the rhetoric of criticality is the question of to what degree the notion of *novelty* in the technical development that is at issue is to be acknowledged or to be disenchanted. It is my impression that to err on the side of the latter is a majority position. While this might be methodologically prudent—exhausting the genealogical paths and counter-paths to the present is, one might argue, doing one's scholarly due diligence—it risks fostering an interpretive predisposition that prematurely frames how unfamiliar phenomena are to be understood, even before analysis begins. To claim to account fully for today's techno-socio-economic-semiotic assemblages by recourse to the history and theory of cybernetics, the persistence of the Cartesian dualism, or theories of governmentality (Amoore et al. 2024; Hayles 2025; Jones 2016), means to leave by the wayside other qualities and aspects that elude these paradigms. The situation of apparent novelty, then, can trigger an almost defiant reflex to declare that, indeed, there is nothing new under the sun. In a polemical psychologization, Matthew Kirschenbaum recently described this tendency toward "historicism" to be "a warm and anodyne blanket. Surely there is precedent and analogy [...] whose appeal lies in the reassurance that our own fraught moment will one day be recollected in tranquility" (Kirschenbaum 2024, 326).

In the current climate, such impulses are not only understandable but in many ways inevitable—when the alternative is the hype and sensationalism crafted to seduce investors and a credulous public; when industry and politics converge to elevate tech CEOs into shadow kings; and when technologies themselves become stages on which power is exercised and enacted at home and abroad, in systems of discipline, surveillance, and war. Yet there is also a danger that this stance, in its reflexive suspicion, ends up explaining away the new by reducing it to an archaeology of preexisting structures, practices, and techniques. In doing so, it risks dulling critique's capacity to confront the present on its own terms by seeking to tame the powers that be through the familiar yet symbolic and in the end mostly comforting gestures of unmasking.

## 3 SURFACE INTEGRITY

There is another way of putting this observation that draws together all three approaches. What the epistemology, the epistēmē, and the epistemics of ML have in common is a shared concern with the hidden, the implicit—with, again citing Philip Agre, what is "underneath." In short, they all operate within a rhetorical framework of *privileging depth over surface*.

In this respect, Critical AI Studies shares a continuity with any humanistic project of critique that operates along a hierarchical distinction between the veiled and the disclosed, the visible and the invisible, the manifest and the latent—from Lacanian psychoanalysis to Derridean deconstruction to Althusserian structuralist Marxism. They all work on the assumption that interpretation must seek, as Frederic Jameson has put it, "a latent meaning behind a manifest one" (Jameson 1994, 60), and that the obvious interpretation of an artifact is obviously wrong. In such "paranoid readings" (Sedgwick 1997), the truth is always hidden behind the surface and needs to be uncovered through a nontrivial hermeneutic effort. We need not linger on the metaphor of the *black box* (Pasquale 2016) here to see that it can equally stand for the operation of the symbolic order, the material practices of ideology, and the opacity of the processes underlying a machine learning model—latent meaning in latent spaces, hidden truths in hidden layers.

It is worth stressing, however, that the logic of suspicion that shapes Critical AI Studies has also a direct continuity with the more media-studies inflected strand of algorithmic critique. This latter approach presents itself as anti-hermeneutical yet continues to privilege depth. Espen Aarseth's classic 1997 work *Cybertext*, for instance, strives to demystify the surface/depth relation by redefining it as purely procedural: "textons" (the stored strings in a system) are transformed by a traversal function of a program into "scriptons" (the strings presented to a reader), so that in computational systems, "the surface of reading [is] divorced from the stored information" (Aarseth 1997, 62–3, 10). In this view, depth is the mechanism of generation rather than a hermeneutical latent meaning, while surface becomes pluralized into the many outputs that can be produced from the same inputs, shifting attention away from hidden truths toward an analysis of how surfaces are technically pro-duced. This move is also evident in N. Katherine Hayles's classic essay "Print is Flat, Code is Deep," in which the distinction between "surface text" and "deeper coding levels" becomes the linchpin for "media-specific analysis" (Hayles 2004, 83, 76). Both Aarseth and Hayles seek a sober, technically informed model of interpretation, yet both retain the hierarchy in which textonic depth take precedence over scriptonic surface, making the former the condition for grasping the latter. Even before the present concentration on machine learning, then, algorithmic critique often hinged on what Nick Seaver has identified as the Aristotelian device of *anagnorisis*, the dramatic revelation of a hidden truth (Seaver 2019).

From both directions, the project of Critical AI might conceivably find itself confronted with the claims of "post-critique." Connected to thinkers such as Eve Kosofsky Sedgwick, Bruno Latour, and Rita Felski, it finds that "critique has run out of steam" (as Dhaliwal quoted Latour above) once its rhetoric of unmasking—long tied to progressive aims—hardens into a mode of conspiratorial thinking that also unmasks, but for decidedly reactionary ends (Latour 2004). Post-critique responds to this situation by reinstating the value of the surface, resisting above all the gesture of declaring X to be, in reality, *nothing but* Y, thus running the risk of missing what might indeed by novel and only speaking to the initiated.[4] That indeed the surface becomes the principal playground for post-critique is exemplified most clearly by Stephen Best and Sharon Marcus's concept of "surface reading," which questions the assumption that depth, *and depth alone*, is worth analysis (Best and S. Marcus 2009). Rather than pursuing a hermeneutics of suspicion, this approach emphasizes the immediacy, materiality, and manifest meaning of an artifact, treating it as a phenomenon to be encountered on its own terms: allowed, as it were, to stand and show itself before one proceeds to unearth, unmask, or debunk what lies beneath.

In methodological debates within Critical AI Studies, it is important to see that these two strands of critique are at play: on the one hand, the hermeneutics of suspicion inherited from literary theory, and on the other, the media-studies mode of algorithmic critique that privileges operational depth. To the first, post-critique and surface reading have already offered responses, calling for renewed attention to what is manifest, immediate, and present at the level of the surface. But their insights have rarely been applied to the second. If extended to algorithmic critique, surface reading would insist that the operational level—code, architectures, or hidden layers—cannot remain the privileged locus of explanation. To treat code as a hidden essence (what Wendy Hui Kyong Chun (2008) has called "sourcery") whose revelation yields truth is to reproduce the very logic of suspicion that post-critique resists. What must be displaced, in both cases, is the staging of critique as revelation of a truth that lies *exclusively* in what is hidden.

The contravening commitment to engaging with what is present, visible, and manifest, rather than subordinating it to a presumed depth, acknowledges what I call "surface integrity." In the context of ML, surface integrity means taking the outputs of models—texts, images, patterns—not *only* as reducible to their underlying technical operations or as symptoms of deeper socio-ideological systems, but *also* as sites of meaning and affect in their own right. I suspect I have to stress this *not only—but also* structure: I am not suggesting setting up a new hierarchy of surface *over* depth—simply reversing the old one, where depth trumps surface—but sidestepping it altogether: Surface and depth demand equal attention. Yet insofar as the surface is concerned, this constitutes a knowledge interest not fully captured by the triad of epistemology, epistēmē, and epistemics I discussed earlier. Such an enterprise is rather, to suggest a name for this, concerned with machine learning's *aisthesis*—in its original sense a type of knowledge acquired through the senses, rooted in appearances.[5]

One reason to follow this track is the simple fact that surfaces are what we primarily engage with in the absence of depth. The fraught rhetoric of the black box notwithstanding (Geitz et al. 2020; Judelson and Dryden 2024), it is certainly correct that ML systems are radically inaccessible for human subjects. I mean this not so much in the sense that there is an explanation gap in retracing the relationship of input and output, but rather that there is a radical inaccessibility of these processes to the imagination—an *intuition gap*. For ML systems are an extreme form of what Edmund Husserl has called "technization," the reduction of meaning-making processes to vicarious symbolic and increasingly autonomous operations that lose their connection to their motivating basis in the life-world of human experience (Husserl 1970, 79). No one can imagine a vector space with thousands of dimensions, and yet we reckon with them to great success; the forward pass of GPT-3, including the simultaneous operation of its 96 attention heads, may be explained in mathematical and computational terms, but it cannot really be intuited. For Husserl, technization is a threat to the very foundations of a notion of knowledge because meaning is generated through procedures inaccessible to intuitive fulfillment, and, more fundamentally, because we rely on them as if they were transparent carriers of the resultant meaning. The danger is that we forget this inaccessibility and thereby allow the autonomy of symbolic operations to eclipse their origin in human experience, replacing the ideal of fulfilled intuition with abstract formalizations and cloaking them in a "garb of ideas" (Husserl 1970, 51). Speaking of an intuition gap in the strict sense suggests that ML models are not just "unfathomable," as a famous paper critiquing the often unknown training sets puts it (Bender, Gebru, et al. 2021, 613). Rather, much of their work belongs to an epistemic regime that is inaccessible to human cognition in principle. In this sense, what Shane Denson calls "discorrelation" is an intensification of technization: the fact that contemporary technical operations fall "out of sync" with the structures of human perception and imagination, producing effects that are real and actionable but not apprehensible (Denson 2020).[6]

[4] Of course, the critique of critique predates Paul Ricoeur's formulation of the "hermeneutics of suspicion." Nor must it be conservative in orientation. One need only recall the mid-twentieth-century debates on secularization, which interpreted modernity as nothing more than Christianity in disguise. In this context, Hans Blumenberg objected that such charges of "historical illegitimacy" presuppose an overly simplistic "unequivocal relation between whence and whither" by variably repurposing the formula " 'B is the secularized A' " (Blumenberg 1983, 4). Already before Blumenberg's 1966 book, Judith N. Shklar had, in a pointed review of Hannah Arendt's *Between Past and Future*, balked at the reductive logic of secularization theses. She noted that those who insisted that the idea of inevitable progress was "really only" secularized Christianity participated in a wave of "debunking" that sought to prove that everything was "really" just like something else—that the great enterprises of the past amounted to nothing more than "small change in an intellectual treasure filled with nothing but copper coins." The result of such analyses, Shklar argued, was that "everything looks 'really' just like everything else and is thus perfectly incomprehensible and pointless" (Shklar 1963, 288).

[5] The term *aisthesis* (αἴσθησις, Greek for perception through the senses) was reintroduced into philosophical discourse by Alexander Gottlieb Baumgarten, who in his *Aesthetica* (1750/58) sought to establish aesthetics as a legitimate form of knowledge. For Baumgarten, *aisthesis* denoted *cognitio sensitiva*, a sensory mode of cognition distinct from but no less valid than conceptual or logical reasoning. What is at stake here, then, is not a diminished or merely preliminary stage of knowledge, but a form of understanding proper to appearances themselves. In this sense, *aisthesis* names a knowledge of the surface that is neither derivative of depth nor reducible to it, but productive in its own right.

The humanities will not be able to hope, as Husserl still did, to restore the operation of abstraction, by way of phenomenology, to presence and fulfilled intuition. Rather, and following his phenomenological critic Hans Blumenberg, it makes more sense to take the negative concept of technization as a loss of presence and turn it into a positive one, a type of technical "unburdening" (*Entlastung*)[7] from the demand of immediacy (Blumenberg 2020). And that means, rather than mourning this presence, to accept that we simply *are* engaged in the use of technical objects, in a praxis that may be more knowing-how than knowing-that, but one that nevertheless still is a type of knowing. From the perspective of a science of culture, this means that technization, realized in technical artifacts, first and foremost brings forth phenomena worth considering in their own right, but also that they are the *conditio sine qua non* of the apprehension of further phenomena. For that reason, they not only integrate into our life-world but become a fundamentally condition for it, and this is what renders them suitable as legitimate objects of investigation—especially as phenomena of *aisthesis*.

Two points seem to follow from this: First, as *surfaces*, these artifacts become part of a world we share with others who perceive and interact with them and it thus constitutes a plane of engagement where meaning is collectively produced. As Gilbert Simondon suggests, technical objects evolve into autonomous concretizations within this common life-world, embedding themselves as active participants in the ongoing production of meaning (Simondon 2017, 1). Such a view opens a path to understanding ML systems not solely as more or less black boxes to pry open to uncover a secret depth, but also as mediating structures that contribute to the ongoing negotiation of cultural, social, and semiotic values—as surfaces we, as a matter of fact, simply are surrounded by. The attention to the surface, then, means sharing a boundary object not only with those who have our intellectual and political commitments in common but with potentially everyone who shares the experience of these surfaces in the first place. The knowledge interest that focuses on the surface is a knowledge interest in *aisthesis* in its communicative dimension. The *Kulturwissenschaften*, to cite Cassirer again, are precisely aimed at this dual aspect: The phenomenon of *aisthesis* they study shows a something, an *aliud* or "it," but also shows this it to a "you," an *alter ego* that is not us but to which we need to refer (Cassirer 1961, 93). The study of culture, then, must be especially attuned to such *aliud/alter ego*-structures of an intersubjectively shared reality. What this means, however, is that taking seriously that surfaces participate in a semiotic process in which appearances become communicative acts may be a way to addresses the very question of audience. As post-critic Rita Felski writes: "Reassessing critique [...] is not an abandonment of social or ethical commitments but a realization [...] that these commitments require us to communicate with intellectual strangers who do not share our assumptions." (Felski 2015, 186) The surface, then, becomes a nexus of communication with others who, indeed, might not immediately be compelled by the well-rehearsed, maybe themselves technicized, gestures of criticality.

The second point follows from recognizing that, despite the apparent abyss between the sciences of nature and of culture, their epistemic regimes nevertheless share a common challenge: how to render intuitive what cannot itself be directly intuited. For the intuition gap is not just an issue for humanists and laypeople, but for the STEM disciplines themselves. Despite Husserl putting the blame on the sciences since Galileo, they have found their own ways of reconnecting formalization with intuition: attention visualizations, saliency maps, and other imaging tools reprise the role of diagrams and illustrations Peter Galison analyzed as attempts to grasp the abstract in the concrete, while at the same time hoping to stave off the imagophobia connected to their all-too literal allure, which he summed up with the constitutive double-bind of all natural science: "We must have images; we cannot have images." (Galison 2002, 300) In this way, as Christina Vagt has discussed recently, the very process of modeling data is an operation of *aisthesis* that produces the knowledge it appears to simply represent—bringing forth the objects of its study in the first place (Vagt 2020).

[6] Recently, Denson has extended this critique to ML models and their latent spaces as structuring and conditioning phenomena while themselves remaining phenomenally inaccessible (Denson 2024).

[7] The term "relief" comes from Arnold Gehlen's anthropology that conceives of technology as a way to make up for the human's lack of natural adaptations (Gehlen 1988).

If intuition plays a role in both, and if the sciences rely on modeling and visualization to meet this challenge, then the question arises: what corresponding techniques are available to the humanities? One answer is to adopt a mixed methodology in which visualization is integrated into humanistic practice, as the digital humanities have especially emphasized, bringing graphical models—maps, networks, embeddings, and other diagrammatic inscriptions—into the reading process not as surrogates for interpretation but as heuristics that render otherwise non-intuitable relations available to judgment, as Sybille Krämer suggests (Krämer 2023). (See Dubreuil (2025) for a critique of this approach) Another path is, again, critique—analyzing the limitations, presuppositions, and blind spots of visualization techniques themselves and emphasizing their constructedness, as Vagt does. Both responses are legitimate and necessary, yet neither addresses the surface integrity I have sought to defend. A third possibility, and one I want to devote the rest of this essay to, would be to cultivate a specifically humanistic *intuition* of machine learning technologies—one that does not begin with depth in order to explain the surface, but that instead takes the surface as its point of departure, and only from there establishes its relation to depth.

## 4 PROBABILISTIC INTUITION, TECHNICAL IMAGINATION

Vilém Flusser—first in his *Communicology* and individual essays and later in the untranslated volume *Lob der Oberflächlichkeit* (*In Praise of Superficiality*, the subtitle promising a "phenomenology of media")—spoke about the necessity of developing a "technical imagination" (*Techno-Imagination*) in response to technical images such as photography, video, and holography (Flusser 1993; Flusser 2022). Technical imagination is Flusser's name for a radical reconfiguration of human thought and social relations brought about both by an increasing loss of meaning and new technological regimes. It denotes the ability to decode and encode these new medial artifacts without falling back on the models of old: "Each code requires a specific method for information manipulation and for deciphering the information coded." (Flusser 2022) Instead of reading a photograph like a painting or a computer image like a photograph, each needs to be understood according to its own inherent logic. One does not have to follow Flusser's ideas about what this logic in each case is in order to realize that he describes a praxis that can read the surfaces of technical structures even in the absence of the impossible fulfilled intention of its underlying processes. To exercise *Techno-Imagination* is the competence of understanding these surfaces on their own terms and to perceive not only the image but the program that makes it possible—the frozen historical and scientific discourses sedimented in the apparatus. And so, lest the surface become detached from its enabling conditions, this means that a technical imagination "requires knowledge of the theories on which apparatuses are based" (Flusser 2012, 198).

In this sense, Flusser very much has an idea of depth in mind, too. But in his phenomenological approach, it does not take explanatory primacy. Instead, Flusser emphasizes negotiating the relationship between both by beginning from the surface rather than the depth, for only the phenomenon of surface can provide the intuition that the abstract knowledge of the technical substructure cannot—which nevertheless one must keep in mind so as to ask the right questions.[8] The challenge for any humanistic inquiry into the surfaces of machine learning outputs today lies in the fact that we have not yet developed the requisite intuition, the technical imagination that brings what we *experience* in relevant contact with how things *work*. Instead, we find ourselves in a collective process of acquiring it—both as professional readers or viewers engaging critically with these outputs and as participants in the culture at large, navigating their pervasive influence—that has its own snags and setbacks.

[8] Here, I find myself in agreement with Shane Denson, who makes a similar case for visual AI as "a transformation registered in subtle modifications of aesthetic categories such as the sublime, the uncanny, and the abject"—provided that his call to "turn from the surface of images to the infrastructures of their generation and processing" is read not as a rejection of surfaces but as an attempt to connect surface and depth, which his phenomenological emphasis strongly suggests (Denson 2023, 151). A fully developed technical imagination would certainly include the task of unlearning "human" seeing and a certain "empathy" for the way machines see as Trevor Paglen has suggested already almost ten years ago. For Paglen, the radical transformation of visual culture lies in the fact that most images today are produced by machines for other machines, operating largely beyond the realm of human perception. To train a technical imagination, then, would mean learning to inhabit this parallel universe of activations, feature maps, eigenfaces, and classifiers—not to naturalize them as "our" way of seeing, but to acknowledge their alterity and the forms of power they enable. Such an effort would not simply transpose human aesthetic categories onto machinic operations; it would also mean developing an intuition for the epistemological and political consequences that follow from the invisibility and automation of vision (Paglen 2016).

Fabian Offert and Ranjodh Singh Dhaliwal have recently formulated a rebuke of the naiveté many humanists bring to machine learning outputs that may count toward a history of this collective learning experience. Their argument also resonates with Flusser's admonition to approach the reading of surfaces with sufficient technical knowledge, to avoid falling into embarrassing errors. In a short intervention entitled "The Method of Critical AI Studies, A Propaedeutic" (Offert and Dhaliwal 2024), they reject a number of "casuistic" humanities responses to ML artifacts that precipitously infer qualities, capacities, and shortcomings of these systems due to a lack of technical understanding. Most interesting in the context of this paper is their discussion of what they call the "benchmark casuistry." It describes the tendency of humanist critics to elevate singular outputs—apparent glitches or just striking examples—into universal judgments about machine learning systems. Put polemically: It is not enough to ask ChatGPT a question, riff on its output, and then call it a day. Offert and Dhaliwal describe such gestures as "n=1" readings: attempts to indict an entire ML system on the basis of one or a handful of outputs, as if the fairness of a die could be judged from a single throw. As they insist, this strategy is fundamentally misaligned with the probabilistic nature of contemporary ML systems, whose outputs vary depending on prompts, contexts, and sampling parameters, and it invites premature triumphalism and false overgeneralization. Unlike deterministic programs of old—the sequential processing of rule-steps—these statistical systems cannot be captured by isolated snapshots or even just a handful of samples.[9] "Probabilistic systems," as the authors put it, "necessarily require a probabilistic form of critique" (Offert and Dhaliwal 2024, 3).

I very much agree with this point. My only addition is this: Just because looking at *one* sample is not enough does not mean that looking at *samples* is wrong. Just because the surface is variable does not mean that no surface is useful—and Offert and Dhaliwal themselves do not make this their claim, even if their preferred remedy is to scale up toward empirical methods and benchmarks across large sets of samples. My suggestion is that this is not the only way forward. What is needed alongside such approaches is the cultivation of a technical imagination by rehearsing a praxis of reading across many surfaces, many system states, and many contexts. Through such interpretive and experiential engagements, one can begin to develop what might be called a probabilistic intuition: a sensibility attuned to the contingent and distributed character of machinic outputs without collapsing them into anecdote.

Such probabilistic intuition is not the reconstruction of the rules that bring about its output. It is itself probabilistic; it cannot grasp the system in its entirety, nor reduce it to a determinate mechanism, but only sketch tendencies, contours, and likelihoods. Such intuition is, in fact, an everyday phenomenon. Fluid dynamics and other complex systems provide instructive examples, precisely because here one can distinguish, at least to a certain degree, between formal technical knowledge and cultivated intuition. Physicists model turbulence or wave formation with equations from statistical mechanics; surfers or sailors develop a practical feel for the sea through long exposure and by learning to anticipate swells or eddies—without solving the Navier-Stokes equations.[10] Both forms of understanding coexist, but they do not stand in for one another. Technical knowledge provides explanatory and predictive power, but it does not replace the embodied, probabilistic sensibility that comes from sustained engagement with the complex phenomenon itself (here, phenomenology and the discourse on qualia converge). The same holds for machine learning systems: knowing how architectures are structured, or how training data is processed, is crucial, yet it does not substitute for the interpretive experience of reading outputs in their variability and distribution. What is needed, then, is a double movement: cultivating surface intuition through repeated encounters, and deepening it with technical knowledge, without mistaking either for being sufficient on its own.[11]

[9] To complicate this statement somewhat: Even deterministic programs require multiple samples to reconstruct their function, since a single output rarely discloses the generative logic. The crucial difference, however, is twofold. First, a deterministic program can, in principle, be reverse engineered in its entirety: once its rules are known, its outputs can be predicted with precision. Second, it is possible outputs are finite and fixed, however large the range may be. By contrast, a probabilistic program cannot be reconstructed in full: its behavior is shaped by statistical distributions rather than explicit rules, and its space of potential outputs is effectively unbounded, with variation that cannot be exhaustively captured. – A simple example of a deterministic text generator is Brion Gysin and Ian Sommerville's *I AM THAT I AM* (first drafted by hand in 1959 and later reproduced with a program Sommerville had written in the mid-1960s). This "permutation poem" exhaustively cycles through a fixed string of words, and its program always yields the same complete sequence: given identical input, the output is entirely predetermined by the algorithm. See Pocknee (2019) for a visual reconstruction of the algorithm. For a discussion of the difference between such "sequential" program and the "connectionist" models of today, see Bajohr (2022a).

[10] I thank Marti Hearst for this example.

[11] This, of course, also happens without an explicit research program (Russell et al. 2025).

## 5 STYLE AS UNITY OF COMPLEXITY

This brings me back to the question of which methods the humanities might contribute as counterparts to the visualization techniques of the sciences. It is here that the subtitle of this essay comes into view. My proposal is that what we do when we cultivate an intuition for complex systems on the surface is above all the apprehension of the *style* of the system's outputs given the same parametric settings.

The choice of the term "style" may seem surprising. One could reasonably object that it is too large and overdetermined a concept, given its rich history and competing definitions, that is moreover difficult to adapt to ML systems. Style has been theorized at the level of *epochs*—in Alois Riegl's *Kunstwollen* or Heinrich Wölfflin's morphological categories—at the level of *group practice*—where, as Dick Hebdige demonstrated, style can function as a subcultural expression of resistance to dominant semiotic codes—and at the level of the *individual*—where thinkers from Georg Simmel to Benedetto Croce linked style to the singularity of artistic expression and individual social differentiation (Croce and Ainslie 1978; Hebdige 1979; Riegl 1992; Simmel 1991; Wölfflin 2015).[12] We simply lack a unified idea of style.[13] And while there is, at least since the late 19$^{th}$ century, a connection between style and statistics in the field of (textual) stylometry, the attempt to list quantitative criteria for stylistic qualities has regularly been criticized on the basis that even the most detailed breakdown of style to measurable features is by necessity reductive; it always only yields a slither of the fullness of the experience of style as a holistic phenomenon.[14] Already in 1959, Michel Riffaterre identified this issue in the gap between the statistical account of linguistic deviation from a norm and its aesthetic effect on the side of the reader; no definition of a style in expressly conceptual terms is able to capture its perceived, its phenomenal meaning (Riffaterre 1959). Finally, applying the concept of style to ML systems faces particular obstacles. If style is defined as the "written individual form of literary *intent*" (Riffaterre 1959, 115), then one is immediately drawn into the contentious debates over whether machine-generated texts can be said to possess meaning or intent at all.[15] The common definition of style as a choice among otherwise equivalent utterances—"a 'that way' which could have been chosen instead of a 'this way' " (Bell 2001, 139)[16]—seems to posit an agent capable of making choices. To speak of style in ML systems therefore requires disentangling these definitions from the level of the system itself (Meier-Vieracker 2025).

I am aware of all these difficulties. Nevertheless, I want to propose that style remains a useful category for a humanistic discussion of ML outputs for two reasons. First, style only becomes perceptible across multiple instances and precisely because it is not expressible in purely conceptual terms denotes an inherently probabilistic mode of perception. Second, it emerges from the cumulative impression formed by such a series, which we can describe as gaining a collective intuition of these impressions.

Style, mood, sensibility—these are fuzzy concepts. A sensibility, Susan Sontag noted, is almost "ineffable" and cannot "be crammed into the mold of a system, or handled with the rough tools of proof." It is perhaps best approached, as she did with "camp," "in the form of jottings rather than an essay—a catalog that discloses coherence without prematurely defining it" (Sontag 1978, 276–7). We may appropriate Sontag's principle of cataloging a shared sensibility for our own purposes of style, then, and say: If isolated sampled output cannot yield immediate insights about the system, a *set* of samples can—but only insofar as this set reveals *a shared unity of complexity* across its probabilistic variations that make them part of *that* set. This unity of complexity is what I want to call "style" here. As Sontag shows with camp, it is never just the property of a single thing but always of a series. It is a purely formal, indeed surface-feature that is shared across all elements of that series without being identical with them. A "style" then would be less the exhaustive identification of a necessary index of characteristics, but a fuzzy identity that shines through what the items on this index have in common; style, as Jeff Dolven writes, "holds things together." Even more, "It makes us see wholes where we might be bewildered by parts" (Dolven 2017, 1). Indeed, it is this holistic quality of style that allows us, following literary theorist and mathematician Peli Grietzer, to conceive of style as the "systemic gestalt" of a series—a pattern that emerges across individual instances while maintaining coherent unity. And it is precisely this "gestalt fluency," as Grietzer argues, that distinguishes probabilistic machine learning systems from deterministic systems, and as such enables the apprehension of collective style in their outputs (Grietzer 2025, 25, 21).[17] Where deterministic systems produce rearrangements of repeatable elements,[18] probabilistic systems generate individual elements that share the same style by corresponding to forms that are "individually complex but collectively simple" (Grietzer 2025, 23).

[12] This is synthesized in Hans Ulrich Gumbrecht's thesis that style serves as the mediation between the claims of subjectivity and reality, which is exacerbated in modernity by the proliferation of multiple worlds and the dissolution of any single, transcendent truth that could serve as a stable reference point for stylistic judgment (Gumbrecht 1986).

[13] It is exactly that conceptual slipperiness that made George Kubler choose the term *form* over that of style: "Style is like a rainbow. [...] We can see it only briefly between while we pause between the sun and the rain." Once we try to "grasp it, as in the work of an individual painter, it dissolves into the farther perspectives of the work of that painter's predecessors or his followers" (Kubler 2008, 118).

[14] While not about stylometrics in particular, an influential argument to that effect is Da (2019).

[15] Standing in for this complex discussion, see the contributions in Kirschenbaum (2023).

[16] This idea of style as choice between equivalent utterances of course goes back to structuralists like Roman Jakobson and Louis Hjelmslev.

This understanding aligns with Best and Marcus's definition of surfaces as "the location of patterns that exist within and *across* texts"—style is this *across*, the repetition and difference that signals collective unity as a single, apprehensible appearance (Best and S. Marcus 2009, 11). None of this means that a single ML model is not capable of producing different styles, which it of course is.[19] But it means that given similar initial conditions, sampled outputs will share a particular style. Here, I believe, one can argue for a praxis of honing a technical imagination that offers a surface-level complement to empirical methods, which it cannot replace but possibly augment. Rather than treating outputs as fixed representations or singular texts, we must learn to perceive them as moments in a broader probabilistic field that becomes unified in its production of a shared gestalt that is present as style—and that we can read off the surface.

[17] For a discussion of neural networks as "quasi-holistic" assemblages that can be understood as gestalt-like, see Bajohr (2021). For more on the history of gestalt theory in the context of ML, see Bajohr (2022b).

[18] Developing an intuition for probabilistic states does of course not exhaust the intuition one develops in interaction with complex systems. The reason for this is that the difficulty of anticipating outputs stems not only from stochastic sampling but also from the sheer scale and complexity of the input–output space, which can make results unpredictable even under deterministic conditions. Furthermore, techniques like reinforcement learning with human feedback further reshape this space: it narrows distributions into more predictable channels while introducing irregular contours that complicate intuitive grasp. I thank Marti Hearst for this point.

[19] As Noam Elcott and Tim Trombley note in their discussion of generative AI "photography," such systems yield a variety of styles, from generic outputs that resemble stock imagery to highly specific renderings produced through fine-tuning on curated archives or simply focused prompting. Nevertheless, all outputs of a system in its "latent specificity" are of that style as a shared unity of complexity (Elcott and Trombley 2025). For a discussion of the fact that the concept of style is itself undergoing a transformation in image models such as Dall-E or mid journey—rendering medium, style, and subject interchangeable, detaching them from historical context, and turning them into ahistorical, monetizable patterns—see Meyer (2023). A further problem arises as stylometric studies of a model's "writing style" often proceed by trying to elicit a so-called default output, which is then assessed through quantitative measures such as adverb frequency, readability, or "involvement." Assuming such a default is the source of Luciano Floridi's optimistic claim that each model carries a distinctive "dataprint" that makes it identifiable (Floridi 2025, 30). The problem is not merely that such stylistic markers can be altered or erased through fine-tuning and related techniques; it arises even earlier, from the difficulty of determining what counts as a model's "default" style in the first place once we take the reality of in-context learning into account: because large language models adapt their outputs to the examples and instructions contained in the prompt, even subtle differences in phrasing can shift the stylistic register of the generated text. As a result, it becomes problematic to speak of a model's "default" style at all, since style may be as much a function of the immediate prompting context as of the model itself.

## 6 SURFACE READING STYLES

What might such a praxis look like that hones a probabilistic intuition by interpreting gestalt features of an output as surface—one that treats the surface not as a mere epiphenomenon of a real and primary depth? I cannot claim that I have a good answer for this; this is where this text is presenting an open-ended question rather than defending results. But I can essay two examples that at least gesture in the direction of the type of reading I imagine and that bring the experience of reading together with an awareness of the underlying system. The first highlights the multiplicity of depths given a single surface; the second shows how arguing from depth alone cannot account for or loses sight of the *aisthesis* it is meant to explain.

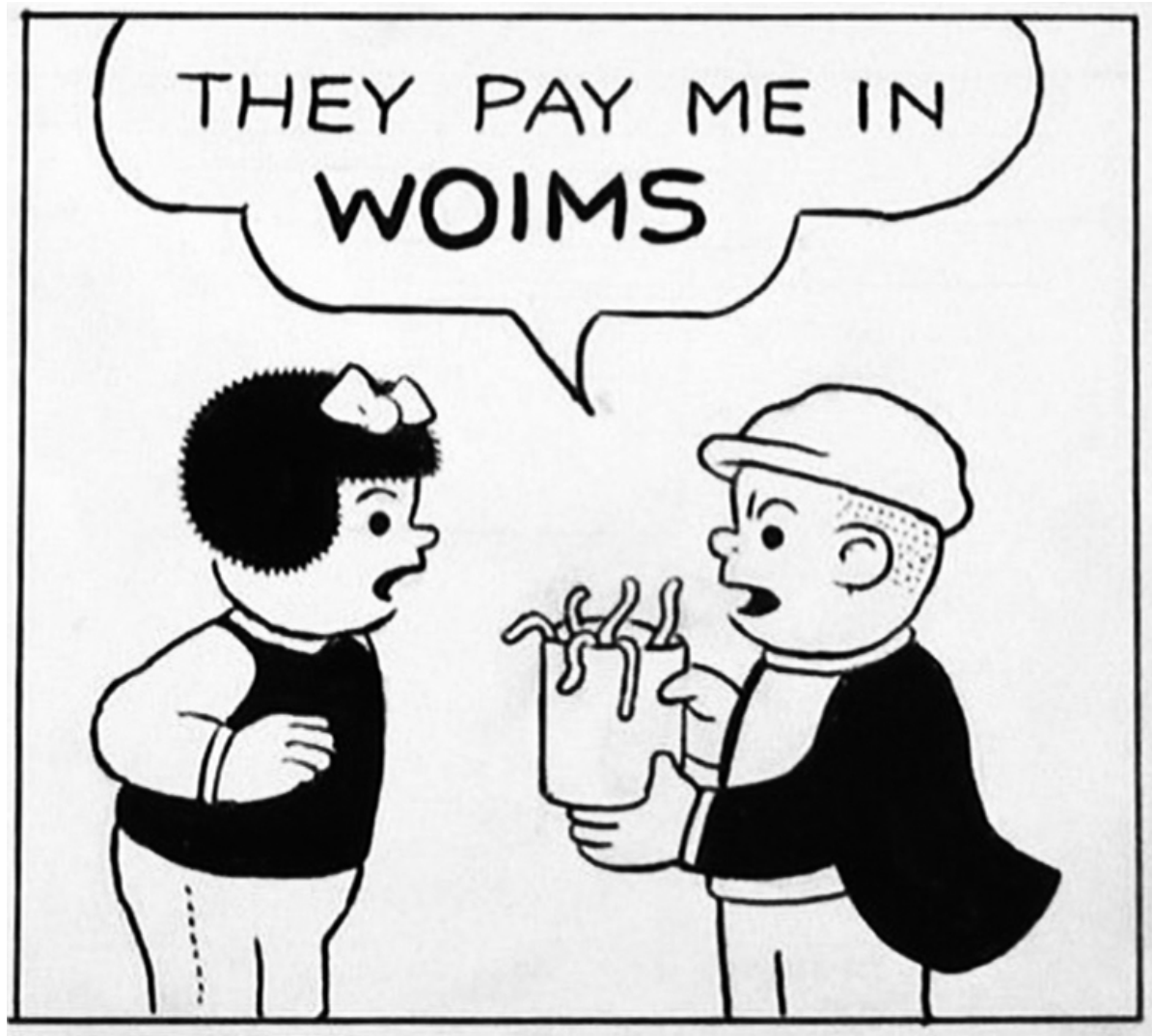

Figure 1: Ernie Bushmiller, detail from a 1978 *Nancy* strip.

### 6.1 Referential Drift

My first example comes from the "wild" of internet communication. In December 2023, Bluesky user Mags Colvett went searching for a *Nancy* comic strip in which the character Sluggo,[20] asked about his new job at a fishing store, holds up a cup of bait and replies in his Bronx accent: "They pay me in woims." The final panel (fig.1), charming in its absurdity, had circulated as a minor meme in 2022. What Colvett found instead was a blog post that we would now call "slop": AI-generated text whose sole purpose is to capture search traffic and generate ad revenue, rather than provide meaningful or accurate content. According to researchers Chantal Shaib et al., its style is characterized by verbosity masking ignorance, low information density, factual unreliability, and tonal incongruity (Shaib et al. 2025).

[20] https://bsky.app/profile/magscolv.bsky.social/post/3kfvud5j5y32s; the post is no longer online.

The text, entitled "They Pay Me in Woims: The Truth about This Bizarre Form of Payment,"[21] is a textbook case of slop thus defined. I want to concentrate on the first aspect of that definition, its verbose unreliability that marks a mode of discourse Gary Marcus and Ernest Davis have already in the age of GPT-3 called "bloviation" (G. Marcus and Davis 2020). The blog post, ostensibly explaining the context and meaning of the panel, offered lengthy and meandering filler that did in fact reveal nothing about the subject.[22] The inability to get to any discernible point is a general quality of slop, but here it was compounded by failing at a most basic linguistic task: resolving "woims" into "worms," a simple act of phonological normalization that most any competent speaker of American English would perform instantly. Instead, the text treats "woims" as a contextless floating signifier that presented a mystery ripe with opportunity for speculation: perhaps it referred to work–life balance; perhaps it was a Reddit in-joke; perhaps a spoonerism of "warm coins." Ironically, it even offered a semiotic interpretation, evoking the role of the reader in the *opera aperta*: "As writer Umberto Eco discussed regarding poetic language, this ambiguity [of meaning] allows readers to contextualize woims based on their own experiences and struggles with work, identity, and purpose."

I want to call the surface phenomenon this text displays "referential drift." The central term, "woims," slips free from its anchoring reference and begins to float, accumulating interpretive possibilities that never coalesce into resolution. The text systematically substitutes metacommentary about meaning for actual semantic closure, delivering a rhetorical performance of knowledge that masks its own epistemic void. This is not mere incoherence but a recognizable style in the sense defined above. A surface reading of slop, properly executed, would tell us how this referential drift appears and is camouflaged on the phenomenon level in some detail.

Such a reading would also connect this phenomenal surface to dimensions of depth—without, however, privileging one as the only one that should count. Following Flusser's principle that technical imagination requires knowledge of the "theories on which apparatuses are based," we can approach this phenomenon through several coordinate frameworks, none epistemologically privileged over the others. One possible context is the technical architecture of language models. The floating reference of "woims" can be traced to tokenization processes and contextual embeddings in transformers. Text is segmented into subword tokens (Shoemaker 2024), then mapped to vectors whose meaning arises through distributional co-occurrence rather than referential grounding—aligning with structuralist intuitions about language as differential relations (Gastaldi 2021). For unknown tokens like "woims," the embedding is determined by subword decomposition and contextual positioning alone. For GPT-4o, for instance, "worms" exists as its own token, while "woims" does not: the tokenizer segments it into known subwords ("wo" + "ims"), each receiving its own vector.[23] The combined representation acquires relational meaning from surrounding tokens and syntactic position, but this meaning simply designates whatever occupies that distributional slot, with no world reference and only sparse relational grounding in token distribution (Mollo and Millière 2025). Confronted with an ungroundable term, the model defaults to generating expansive discourse on ambiguity and meaning-making—a pattern abundant in its training data—rather than performing the pragmatic resolution a human speaker would.[24]

But this technical explanation is only one lens among several. An equally productive approach, and one more in line with the core of Critical AI Studies, would take up the political economy of content farms and search engine optimization (SEO) as a theory "on which apparatuses are based" on par with the technical papers that make them work. The "woims" text exists within an attention economy where visibility trumps accuracy and labor conditions incentivize volume over quality or veracity. Picking up a genre of text in existence before AI—human content mills have long produced bloviation for SEO purposes—the text fills in a template into which referential drift slots seamlessly. Strategic verbosity in both cases is not merely a technical artifact but an economic strategy: more text means more keywords, scroll time, and ad impressions. "Woims" becomes synecdoche for slop itself: a token misrecognized and ungrounded, yet endlessly recycled and monetized precisely because its irresolution permits infinite elaboration.

[21] https://www.ownyourownfuture.org/they-pay-me-in-woims. Unfortunately, the text is no longer online in its original form; the post has been updated with a new title, "Understanding the Enigma: A Comprehensive Analysis of 'They Pay Me in Woims' ", see below.

[22] The term "bloviation," as well as the connected term "bullshit"—understood by Harry G. Frankfurt as speech for which truth is irrelevant—can strictly speaking only be used in the meaning of a reception phenomenon since they imply intent; a similar problem as above with style as choice (Frankfurt 2005).

[23] https://platform.openai.com/tokenizer.

[24] Critics of LLMs highlight this issue as a sign that they cannot truly "understand" language, underscoring the conceptual divide between the appearance of meaningful output and the absence of any genuine semantic foundation in these models, e.g. Bender and Koller (2020). In contrast, the semiotic perspective, rooted in structuralism, views language as a system of signs that derive meaning through their differential relationships to one another rather than through external reference. When applied to LLMs, this semiotic framework suggests that these models succeed not by grounding symbols in the world but by generating language through intricate patterns of internal coherence; reference, in this reading, is but a secondary effect, not a foundational quality of language, e.g. Weatherby (2025). This correlation-based semantics produces what I have elsewhere called "dumb meaning"—not empty, but limited to distributional associations without causal-indexical reference (Bajohr 2024b).

Both the technical as well as the economic interpretation—and there would be more, such as genre theory, that straddles both—illuminate the surface of the "woims" phenomenon differently, yet none is foundational. What is more, both must come *after* the discussion of the surface, which is the prior phenomenon in each case.

### 6.2 Surface Narration

Or so it seems. For when it comes to the technical underpinning, but also the assumption that this technology is used to serve capital interest, "woims" is but fragile evidence and a reminder how high the bar is for a surface reading properly executed. We do not know which model generated it, to which degree it was edited, and our suspicion of it as synthetic text rests largely on stylistic recognition alone—on its appearance-as-slop. To make matters worse, the blog post was at some point updated (though, curiously, by no means made more coherent) so that even the initial text is no longer retrievable. And what is more, most any state-of-the-art LLM *would* indeed understand "worms" to be related to "woims" as a type of misspelling, and likely even GPT-3, the most advanced model at the time the article was published in 2022. Reading ML-generated text as surface, then, also shows some of the difficulties we face in attempting to build a praxis of technological imagination as probabilistic intuition in the absence of knowing for sure that it is indeed synthetic text we are reading.

In the absence of this certainty, we are simply in the same situation as everyone else. For ML-generated text itself is a category imbued with a doubt that stages a "paranoid reading." This is in line with what I have elsewhere described as the "post-artificial" situation: As soon as human-sounding text can be mass-produced, we can no longer be sure of the origin of a text, and doubt becomes a permanent hermeneutic situation (Bajohr 2024c). It generalizes what Gabriel de Seta has called "algorithmic folklore," that is, the folk theories users develop when faced with inscrutable algorithms, from "shadow banning" on social media to the idea that there is a secret language embedded in the outputs of image generators (Seta 2024; Shoemaker 2024). A related phenomenon is the more recent heuristics of what one might call "folk forensics," the claims to revealing AI authorship by identifying markers such as em-dashes, bullet points, or oddly timed punchlines. Folk forensics may have some grounding in a real recognition of ML-generated text, but it can quickly become paranoid too, since, as Shaib et al. note, "text can be perceived as 'slop' even when not generated by AI, and not all AI-generated text reads as 'slop' " (Shaib et al. 2025, 1).[25] In this sense, the case of "woims" did indeed capture a particular collective reading experience, yet it failed to meet the need to verify the conditions of its production, which, while they may not belong to the explanatory dimension of depth, must nonetheless be known to ensure that what we are encountering is in fact AI-generated text.

To remedy this limitation, I want to offer a second example in which surface integrity is preserved but the depth of the model is still present as ascertainable, because—*verum ipsum factum*—I made it myself. It serves to show, too, that arguments drawn from depth alone often miss the phenomena they purport to illuminate (the following repeats points made in Bajohr (2024a)).

Remarking on the capacity of ML systems to create narrative, Angus Fletcher has put forward the argument that any statistical machine learning system is in principle unable to narrate because narration is dependent on structures of causality; causality cannot be properly reproduced via statistical correlations; therefore, LLMs cannot narrate (Fletcher 2021). This, too, is an argument from depth: Fletcher grounds his claim not only in the technical architecture of statistical machine learning but also in its underlying metaphysical assumptions (in the analytic sense of the term). For he never examines actually existing synthetic text and its surfaces. My suspicion is that a surface reading of LLM narrative would show that this inference is not necessary. Such a reading might even allow reasoned insight into the system's capacities that are not reducible to its mere technical substratum. Surface reading LLMs does not simply start with a technical assumption of an inability for causal reasoning, that is, an argument from depth; and it engages in some kind of praxis that hones a technical imagination, a probabilistic intuition, while at the same time making sure we are not just looking at isolated samples.

To make this case, I can draw on my dual identity as both a scholar and an author of digital writing, by pointing to a novel I wrote with an LLM myself. By producing the text to be read in the first place, I was— keeping Flusser's warning in mind—able to stay aware of the underlying technical structures that might be more difficult to come by in a text from a third party, while the repeated process of text generation produced a large enough sample size from the same model that it allowed me to draw inferences about its overarching style. Such a process is, I am aware, not an experiment in any empirically reliable sense, which would allow for replicability and provide objective criteria of evaluation; it is, for lack of a better word, artistic research. And yet this is in keeping with my argument that the approach to developing a probabilistic intuition and exploring gestalt features of LLM outputs can be done engaging experientially. The result, one might add, is the creation of a surface that is open to be engaged with by anyone—a novel. Artistic research as a praxis, I think, aligns well with the idea of *aisthesis* as a legitimate field of collective knowledge-production.

[25] The collective reading of the "woims" blog entry that happened on Reddit reached such a vertiginous point when a user started to doubt whether the image itself was real rather than AI-generated, https://www.reddit.com/r/comicstriphistory/comments/v5gdpb/why_is_the_ernie_bushmiller_comic_strip_nancy_so/.

The result of this praxis was the novel *(Berlin, Miami)* that appeared in German in 2023. It was made using the open LLM GPT-J developed by EleutherAI, fine-tuning its standard model on a set of four German-language novels. The output text is difficult to summarize precisely because it *struggles* with narrativity. At first glance, it seems to possess the elements of a conventional story: recurring characters—such as the mysterious, Odradek-like "Jawling"—unfolding events—such as Miami's succumbing to the sinister "Life Viruses"—snippets of dialogue, and even formal elements like footnotes or transitional passages. The text offers enough to feel as though a story is taking shape, yet a closer look reveals an odd, anti-narrative quality of logical inconsistencies and temporal leaps. Events happen, but their connections are tenuous, even contradictory. Characters appear and reappear, yet their motivations remain opaque. The elements of the story itself are effects of the training set I used, which mostly comprises conventionally narrated novels set in a near, technology-dominated future. Looking past these elements, however, one can try to analyze the style of narration, based purely on the surface of the text as a linguistic artifact, as a collective pattern across many samples. This means foregoing categories such as correlation and causation—as qualities inherent to the depth structure of the technical system producing the text—and instead turning to stylistic features manifest on the surface of the text, such as coherence and cohesion.

If cohesion refers to the way in which text elements are linked at the phonological, orthographic, and lexicogrammatical levels, coherence refers to their abstract meaning context. The separation is of course ideal-typical, so that in any concrete work there is always a mixture of cohesion and coherence. All texts are a combination of either, and each communicates coherence via cohesion, and requires coherence for the cohesion to communicate something. Between the ideal-typical extremes of pure material cohesion and pure ideal coherence as well as the unmarked, conventional texts that balance both, there exists a type of text in which coherence is reflected on the level of cohesion; I believe these are narrative texts. In them, the function of coherence is elevated to the organizing principle of the text, manifested as aggregations of "therefore." In literary narratives, "therefore" functions not only as a marker of logical consequence but also as a structuring element that order story elements into a meaningful sequence of events across space and time. *(Berlin, Miami)* forgoes both the balance of coherence and cohesion that characterizes conventional texts, and the projection of coherence into the organizing principle of narrative. The text is marked by an overabundance of cohesion—its sentences and paragraphs flow smoothly; its recurring motifs create a sense of familiarity—but a lack of consistently generating coherence. The result is a text that simulates isolated patterns of conventional narrative without fully realizing its continuous structure.

From here, one may of course go on connecting this to the underlying technical system, trying to explain it with the lack of causal reasoning or any other property of machine learning in the vein I did above for the "woims" example. Unlike Fletcher's, however, such a reading starts from the surface integrity of its object and will thus not be satisfied with a precipitous transition to a depth structure argued purely on principle. For while the lack of coherence reduces the conventional narrative qualities of the text, such a reading would also acknowledge that the text is not a non-narrative, either. A phenomenologically informed reading would insist that the act of reading itself—the reader's process of sense-making, of *semiosis*—is constitutive of what it means to engage with a surface. Even the objective absence of cohesive or, on a technical level, causal inference can appear as narrative, since the reader's experience necessarily fills in the blanks, the "points of indeterminacy" (Ingarden) or *Leerstellen* (Iser), through the intentional act of reading. Yet this very dynamic constitution of the text on the part of the reader, between its perceived form and the horizon of expectation it elicits, is always a function of the overall gestalt of the text itself—it is a process of becoming accustomed to what range of possible expectations a text allows for. And it is this process of convergence—the gradual alignments of the horizons of the expected, the conventional, and the realized form of the text—where the development of a probabilistic intuition might take place; one, to be sure, that is trained only on *this* text, as samples from *this* model—but an intuition nonetheless.

## 7 CONCLUSION

The surfaces of machine learning outputs are not mere epiphenomena awaiting excavation, but the primary plane on which these systems participate in culture. This essay has argued for taking seriously what I have called "surface integrity"—the recognition that ML-generated texts, images, and patterns possess meaning and affect in their own right, irreducible to either their technical substrates or the socio-political conditions of their production. This is not to abandon depth, but to refuse its explanatory monopoly. The technical architectures, historical genealogies, and material critiques that constitute the depth of ML systems remain indispensable. Yet they cannot fully account for how these systems actually manifest in the life-world, how their outputs circulate as cultural artifacts, or how we collectively learn to make sense of them.

In arguing for a phenomenological approach to synthetic text—one that takes the *aisthesis* of machine learning outputs seriously as a function of their surface integrity—I have taken up some arguments from post-critique that concern questions of audience and novelty, which I believe Critical AI studies must be prepared to confront. I am well aware that a call for surface reading and post-critique, at this historical-political juncture, might sound, at the very least, tone-deaf. Isn't critique the most urgent task? Yet it is precisely for this reason that it may be worthwhile to expand the interpretive scope of the project of critique to encompass all aspects of ML as cultural artifacts, including their *aisthesis*, as a legitimate knowledge interest. The fact that these artifacts exhibit a unity of complexity—manifest in the consistent style across multiple outputs of the same system—demonstrates that they can be read meaningfully.

To be sure, this essay has sketched only the beginning of such a practice. The two examples offered—the referential drift of "woims" and the surface narrative of *(Berlin, Miami)*—suggest directions rather than destinations. A fully developed practice of surface reading LLMs would require more systematic engagement across different models, contexts, and genres; more careful attention to the feedback loops between human readers and machine outputs; more nuanced theorization of how style operates in probabilistic systems. It would also need to remain vigilant against its own potential reifications—resisting the temptation to treat "ML style" as a fixed category rather than an evolving phenomenon shaped by technical development, economic pressures, and cultural reception. To reiterate: The form of surface reading I have present here is meant not to replace but to complement existing critiques of ML systems by foregrounding their experiential and phenomenological dimensions, all the while honing a technological imagination that understands its object not as a singular, deterministic specimen, but as probabilistic sets. In the end, this amounts to not reducing but indeed multiplying the methodologies at our disposal—going both ways, from depth to surface, and from surface to depth, so that, somewhere in the middle, knowledge may appear.